\documentclass[twocolumn,aps,nofootinbib]{revtex4}
\usepackage{amsfonts, amssymb}
\usepackage{graphicx, epsfig,bm}
\usepackage{color}

\newcommand{\be}{\begin{equation}}
\newcommand{\ee}{\end{equation}}
\newcommand{\bea}{\begin{eqnarray}}
\newcommand{\eea}{\end{eqnarray}}

\def\fun#1#2{\lower3.6pt\vbox{\baselineskip0pt\lineskip.9pt
        \ialign{$\mathsurround=0pt#1\hfill##\hfil$\crcr#2\crcr\sim\crcr}}}

\newcommand\lsim{\mathrel{\rlap{\lower4pt\hbox{\hskip1pt$\sim$}}
    \raise1pt\hbox{$<$}}}
\newcommand\gsim{\mathrel{\rlap{\lower4pt\hbox{\hskip1pt$\sim$}}
    \raise1pt\hbox{$>$}}}

\def\dslash{\not{\hbox{\kern-2pt $\partial$}}}
\def\Dslash{\not{\hbox{\kern-4pt $D$}}}
\def\Oslash{\not{\hbox{\kern-4pt $O$}}}
\def\Qslash{\not{\hbox{\kern-4pt $Q$}}}
\def\pslash{\not{\hbox{\kern-2.3pt $p$}}}
\def\kslash{\not{\hbox{\kern-2.3pt $k$}}}
\def\qslash{\not{\hbox{\kern-2.3pt $q$}}}

 \newtoks\slashfraction
 \slashfraction={.13}
 \def\slash#1{\setbox0\hbox{$ #1 $}
 \setbox0\hbox to \the\slashfraction\wd0{\hss \box0}/\box0 }

\def\ee{\end{equation}}
\def\be{\begin{equation}}

\begin{document}
\setlength{\unitlength}{1mm}
%\twocolumn[\hsize\textwidth\columnwidth\hsize\csname@twocolumnfalse\endcsname]
\title{Lensed Cosmic Microwave Background Constraints on
Post-General Relativity Parameters}
\author{Paolo Serra$^1$, Asantha Cooray$^1$ Scott F. Daniel$^2$, Robert Caldwell$^2$, Alessandro Melchiorri$^3$}
\affiliation{$^1$Center for Cosmology, Department of Physics and Astronomy,
University of California, Irvine,  CA 92697}
\affiliation{$^2$Department of Physics and Astronomy, Dartmouth College,
Hanover, NH 03755 USA}
\affiliation{$^3$Physics Department and Sezione INFN, University of Rome,
``La Sapienza,'' P.le Aldo Moro 2, 00185 Rome, Italy}
 
\date{\today}%
%\maketitle % use with old revtex !

\begin{abstract} 

The constraints on departures from general relativity (GR) at cosmological
length scales due to cosmic microwave background (CMB) data are discussed. 
The departure from GR is measured by the ratio, 
parameterized as $1 +\varpi_0 (1 + z )^{-S}$, between the gravitational potentials conventionally appearing in the geodesic equation and the Poisson equation. 
%Naive arguments motivated by the accelerated cosmic expansion
%suggest $\varpi_0$ of order unity and $S \sim 3$, in which case the departure
%from GR becomes significant at late times. 
Current CMB data indicate $\varpi_0 =
1.67^{+3.07}_{-1.87}$ at the 2$\sigma$ confidence level, while $S$ remains unconstrained. 
The departure from GR affects the lensing conversion of E-mode into B-mode polarization.
Hence, the lensing measurements from a future CMBpol 
experiment should be able to improve the constraints to $\varpi_0
< 0.30$ for a fiducial $\varpi_0=0$ model and independent of $S$. 

\end{abstract}
\bigskip

\maketitle
%$\varpi_0 < 4.99$ and $S > 1.97$
%%%%%%%%%%%%%%%%%%%%%%%%%%%%%%%%%%%%%%%%%%%%%%%%%%%%%%%%%%%%%%%%%%%%%%%%%%%%%%%%
%\section{Introduction}
%%%%%%%%%%%%%%%%%%%%%%%%%%%%%%%%%%%%%%%%%%%%%%%%%%%%%%%%%%%%%%%%%%%%%%%%%
%%%%%%%%%%%%%%%%%%%%%%%%%%%%%%%%%%%%%%%%%%%%%%%%%%%%%%%%%%%%%%%%%%%%%%%%%

\noindent \emph{Introduction---}The quest for the source of the cosmic
acceleration has led to speculation that the proper theory for gravitation
departs from general relativity (GR) on cosmological length scales (e.g.
Ref.~\cite{Uzan:2006mf}). There are numerous theoretical examples that introduce
new gravitational degrees of freedom and that are capable of producing a
late-time acceleration, with wide-ranging implications for observable phenomena
(e.g. Refs.~\cite{Dvali:2000hr,Carroll:2003wy}). Given this possible abundance in
new physics, it is important to identify tests that can distinguish between the
effects of dark energy and those of modified gravity. Though late-time
accelerated cosmic expansion is the principal indicator that a new ``dark''
physics is needed, it is not the only test such physics must satisfy. A
successful cosmology must also agree with measurements related to the behavior of
inhomogeneities as probed by the cosmic microwave background and large-scale
structure.

To understand the extent to which cosmological data support GR, we make use of an
approach motivated by the post-Newtonian parameterization of the gravitational
field within the Solar system and introduce a post-GR parameterization for
cosmological perturbations. Such a parameterization is also motivated by the
common feature within a broad range of gravity theories of a decoupling of the
perturbed Newtonian-gauge gravitational potentials $\phi$ and $\psi$, defined by
the perturbed Robertson-Walker line-element
\begin{equation}
\label{metric}
ds^2=a^2 \left[-\left(1+2\psi\right)d\tau^2
+\left(1-2\phi\right)d\vec{x}^2\right] \, ,
\end{equation}
using the notation and convention of Ref.~\cite{Ma:1995ey}. 
%To give a physical
%sense of these potentials, $\psi$ enters the Newtonian limit of the equation of
%motion, $\ddot{\vec x} = -\vec\nabla\psi$, and $\phi$ enters the Poisson
%equation, $\nabla^2\phi=-4\pi G a^2\delta\rho$.

Whereas GR predicts $\psi=\phi$ in the presence of non-relativistic matter,  a
{\it gravitational slip}, defined as $\psi\neq\phi$, occurs in
modified gravity theories. For example, this inequality means that the
gravitational potential of a galaxy cluster is not the same potential
traced by the geodesic motion of the constituent galaxies. 
Hence, a new relation between these potentials is a launching point for investigations of cosmological
manifestations of modified gravity \cite{Hu:2007pj,Bertschinger:2008zb}. 
For primordial cosmological perturbations, the potentials are not completely free, however, as
there exists a constraint equation in the
long-wavelength limit \cite{Bertschinger:2006aw}. 

\begin{figure}[!t]
  \begin{center}
    \begin{tabular}{ccc}
 \resizebox{80mm}{!}{\includegraphics{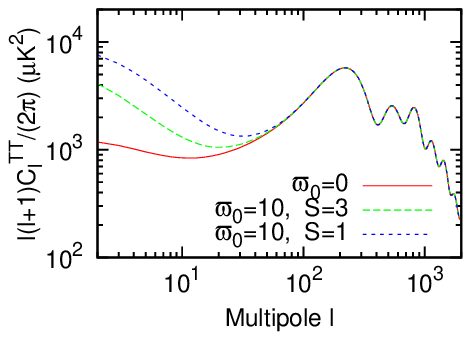}} \\
  \resizebox{80mm}{!}{\includegraphics{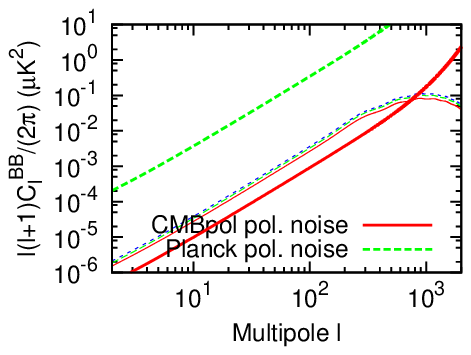}} \\
   \resizebox{80mm}{!}{\includegraphics{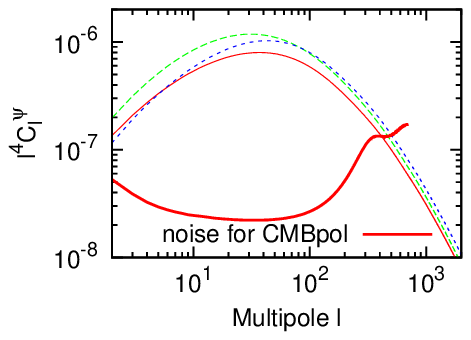}}  \\
    \end{tabular}
    \caption{The effect of the modified gravity parameters on the temperature,
B-mode polarization, and lensing potential  power spectra for the best fit
$\Lambda$CDM model from WMAP5+ACBAR \cite{Komatsu:2008hk}; we also show the measurement
error for B-mode polarization measurements with Planck and CMBpol 
and the CMBpol noise for lensing reconstruction (up to $l=700$).}
    \label{fig1}
  \end{center}
\end{figure}

We consider an alternative theory of gravitation that predicts an expansion
history indistinguishable from $\Lambda$CDM, accompanied by post-GR effects
whereby 
\begin{equation}
\psi(\tau,\,\vec x) = \left[1+\varpi(\tau,x) \right] \times
\phi(\tau,\, \vec x), \label{eqn:varpi}
\end{equation} 
following Refs.~\cite{Caldwell:2007cw,Daniel:2008et}. If the new gravitational
phenomena is to mimic the effects of $\Lambda$ by changing the amount of
spacetime curvature produced by the cosmic matter density, then we expect
$\varpi$ to grow to order unity at late times on large scales. Looking for clues
to such a scenario, CMB temperature anisotropies alone provide a weak constraint to $\varpi$ as the
departure from GR is primarily manifest in the integrated Sachs-Wolfe effect
\cite{Lue:2003ky,Hu:2008zd}, as illustrated in Fig.~\ref{fig1}. However, CMB lensing
is also sensitive to $\varpi$ because the lensing deflection of CMB photons by foreground large-scale 
structure depends on
the sum of the potentials $\psi+\phi$
\cite{Huterer:2006mva,Acquaviva:2004fv,Schimd:2004nq,Calabrese:2008rt}. In this {\it Letter}, we
show that the expected conversion of E-mode to B-mode polarization through
lensing \cite{Zaldarriaga:1998ar}, shown in Fig.~\ref{fig1}, allows a new probe
of departures from GR that will be accessible to future CMB B-mode polarization
experiments.

The lensing of the CMB affects temperature perturbations at the level of a few
percent at arcminute angular scales, which is on the damping tail of CMB
anisotropies \cite{Lewis:2006fu}. Using temperature anisotropy data from WMAP
\cite{Komatsu:2008hk} and ACBAR
\cite{Reichardt:2008ay} we can only put weak constraints on the  post-GR
parameterization at present. On the other hand, B-modes at
tens of arcminute angular scales are mainly due to the lensing
conversion from E-modes.   Using the combination of E- and B-modes
one can reconstruct the lensing signal in CMB data by using quadratic statistics
\cite{Hu:2001kj,Cooray:2002py,Kesden:2003cc} and likelihood methods
\cite{Hirata:2002jy}. The projected lensing potential power spectrum out to
the last scattering surface can then be used to extract $\varpi$. 
As we find, upcoming high sensitivity CMB polarization experiments, 
such as CMBpol \cite{Baumann:2008aj,Bock:2008ww}  of
NASA's Beyond Einstein program, have a significant role to play in
constraining GR at cosmological length scales.

\smallskip
\noindent \emph{Calculational Method---}The treatment of cosmological
perturbations under modified GR follows from Ref.~\cite{Daniel:2008et}. 
The metric perturbation variables in the
synchronous and conformal Newtonian gauges are related as
$\psi = \dot\alpha + \mathcal{H}\alpha$, 
$\phi = \eta - \mathcal{H}\alpha$,
where $\alpha \equiv (\dot h + 6 \dot\eta)/2 k^2$,
$h,\,\eta$ are synchronous-gauge metric variables, and
the dot indicates the derivative with respect to the conformal time \cite{Ma:1995ey}. 
In GR ($\varpi=0$), the perturbed Einstein equations, 
\begin{eqnarray}
k^2\eta-\frac{1}{2}\mathcal{H}\dot{h}&=&4\pi Ga^2\delta T^0_0\label{zerozero}\\
k^2\dot{\eta}&=&4\pi G a^2 (\bar{\rho}+\bar{p}) \theta\label{zeroi}\\
\ddot{h}+2\mathcal{H}\dot{h}-2k^2\eta&=&-8\pi G a^2\delta T^i_i \, ,\label{ii}
\end{eqnarray}
are used to evolve the metric variables, where $(\bar{\rho}+\bar{p})\theta\equiv
i k^j\delta T^0_j$ \cite{Ma:1995ey}. 

In our post-GR description, we assume the stress-energy tensor is conserved and
that there is no preferred reference frame introduced by the new gravitational
effects. Consequently,  Eq.~(\ref{zeroi}) remains valid but
Eqns.~(\ref{zerozero},\,\ref{ii}) do not. 
%
% In fact, if the perturbed Einstein equations were assumed to remain valid, then
% the gravitational slip would necessarily imply the existence of new energy
% density and pressure perturbations which are comoving with the baryonic and dark
% matter density perturbations.
%
Because gravitational slip is degenerate with a cosmological fluid component with
shear, Eq.~(\ref{eqn:varpi}) becomes
\begin{equation}
\dot{\alpha}=-(2+\varpi)\mathcal{H}\alpha+(1+\varpi)\eta  
 -12\pi G a^2(\bar{\rho}+\bar{p})\sigma/k^2. \label{alphadot}
\end{equation}
This modification preserves the
consistency condition for long wavelength cosmological perturbations 
\cite{Bertschinger:2006aw,Daniel:2008et}.

In our study we restrict attention to a homogeneous model of gravitational slip,
\begin{equation}
\varpi = \varpi_0 (1+z)^{-S},
\end{equation} 
and seek to constrain the post-GR parameters
$\varpi_0$ and $S$. By allowing the redshift dependence of the modified gravity
parameter $\varpi$ to be a free parameter, this relationship is more general than
the one introduced in Ref.~\cite{Caldwell:2007cw}. (Note that we have also
removed a prefactor $\Omega_\Lambda/\Omega_m$.) 

The lensing deflection of CMB photons by foreground large-scale structure depends
upon gradients in the total gravitational potential $\phi+\psi$ transverse to the
line of sight to the last scattering surface \cite{Lewis:2006fu}. 
The evolution of the  lensing potential is separated from the 
primordial curvature perturbation $R(\vec{k})$ using a transfer function
$T_{\phi}(\vec{k},\tau)$, whereby
$\phi(\vec{k},\tau)=T_{\phi}(k,\tau)R(\vec{k})$.  Hence, the power spectrum of
the lensing  potential is
\begin{equation} 
C_\ell^{\phi}=4\pi\int\frac{dk}{k}P_R(k)
\Big[\int_0^{\chi_*}d\chi\,S_{\phi}(k;\tau_0-\chi)j_\ell(k\chi)\Big]^2 \, .
\end{equation}
Here $P_R(k)$ is the primordial power spectrum, $\tau_0-\chi$ is the conformal
time at which a given photon was at the position $\chi\hat{n}$, and the lensing
source is given by:
\begin{equation}
S_{\phi}(k,\tau_0-\chi)=(2+\varpi)T_{\phi}(k,\tau_0-\chi)
j_\ell(k\chi)\Big(\frac{\chi_*-\chi}{\chi_*\chi}\Big) \, ,
\end{equation}  
where we have made use of the post-GR relation between $\phi$ and $\psi$ to
simplify the expression in terms of the transfer function of $\phi$. To evaluate
the lensing source and angular power spectrum, we use Eqns.~(\ref{zeroi},
\ref{alphadot}) to evolve $\eta$ and $\alpha$, from which $\phi$ is obtained.
The lensing potential for different values of the post-GR parameters is shown in
Fig.~\ref{fig1}.  In the case of temperature, lensing modifies the damping tail.
The B-mode polarization signal due to lensing  that peak at tens of arcminute angular scales is
directly proportional to the lensing power spectrum. 
We ignore non-linear corrections to the lensing calculation 
as non-linearities are responsible for less than a 6\% change to the B-modes \cite{Lewis:2006fu}
and we only consider parameter constraints out to $l < 700$ when using $C_\ell^{\phi}$.

\begin{figure}[!t]
\begin{center}
\begin{tabular}{cc}
\resizebox{70mm}{!}{\includegraphics{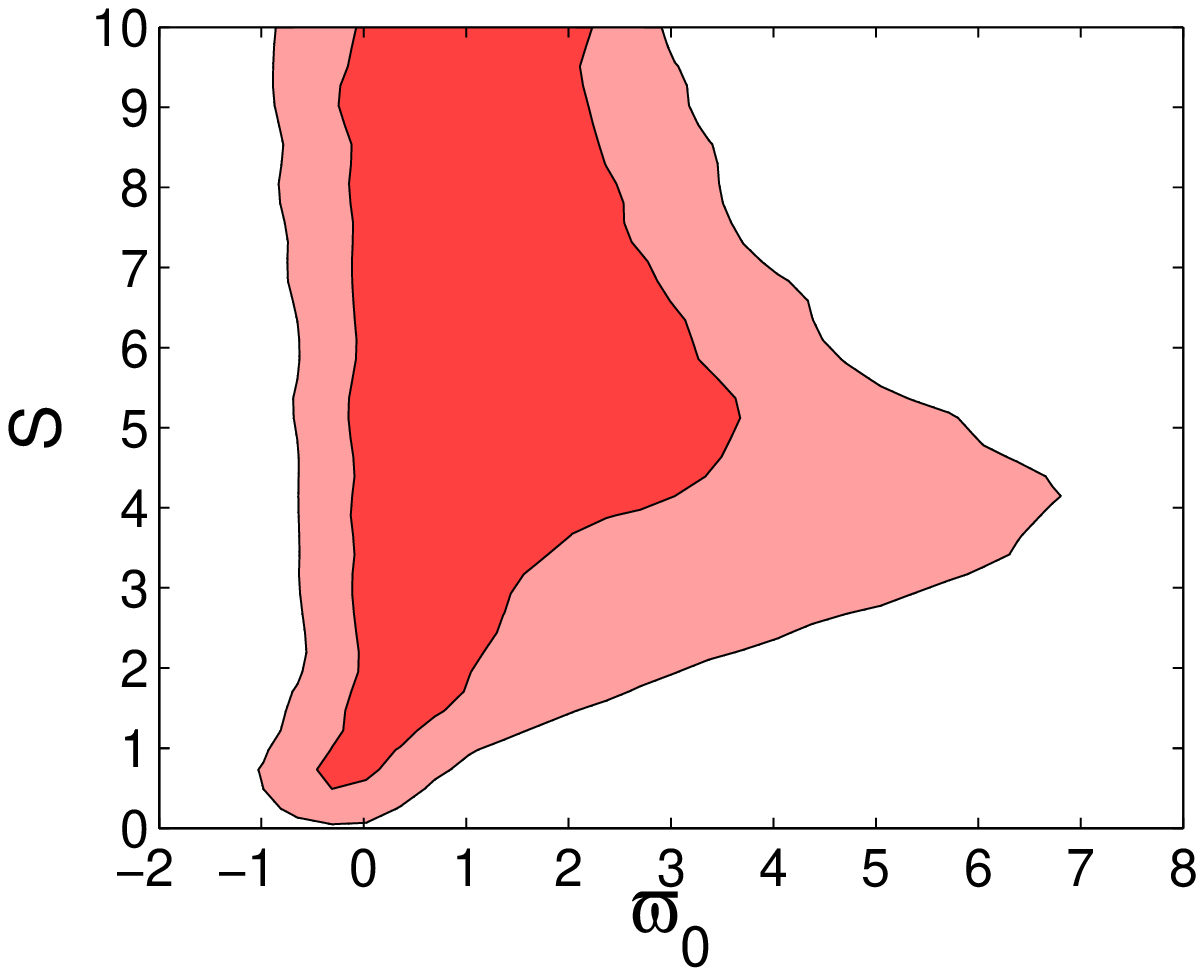}} \\
  \resizebox{70mm}{!}{\includegraphics{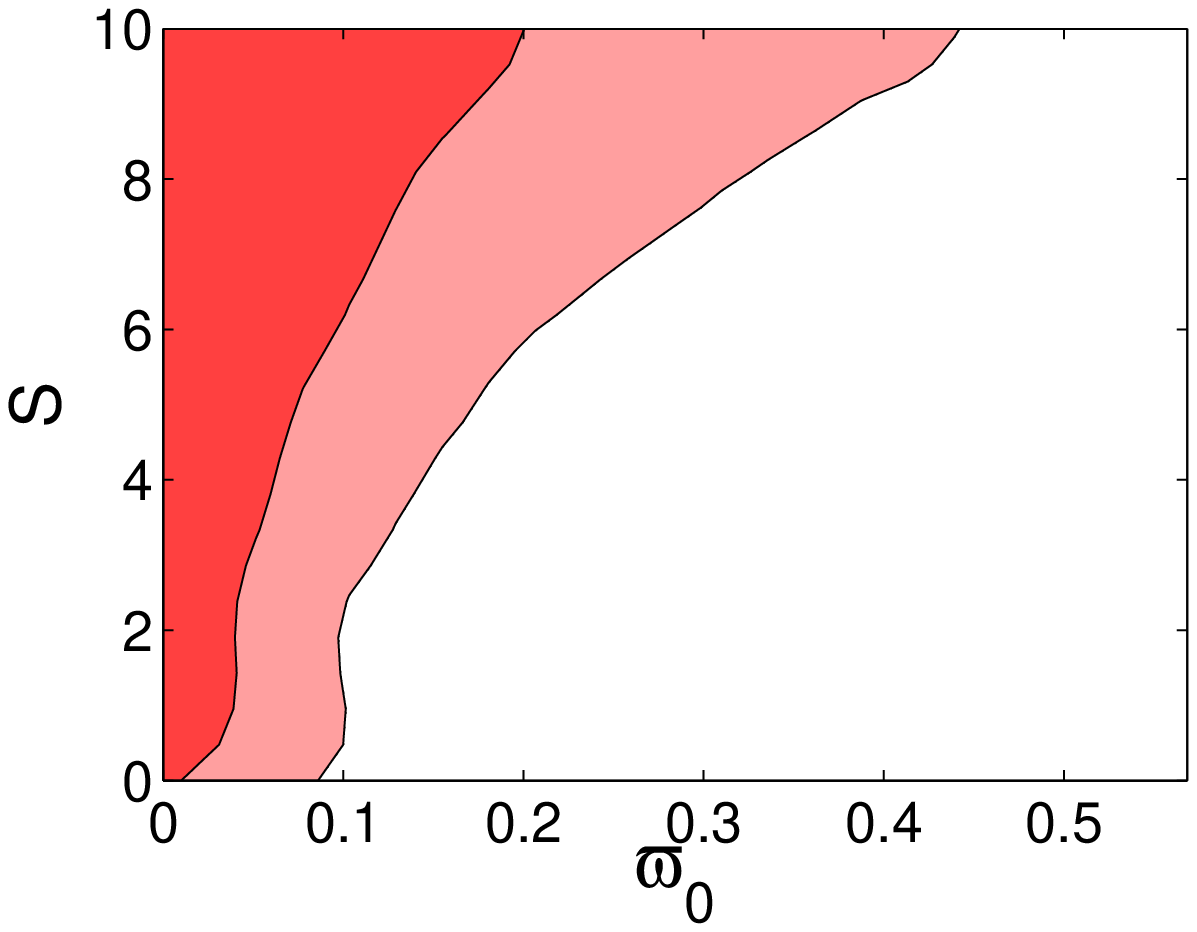}} \\
\end{tabular}
\caption{Two-dimensional contours at 1$\sigma$ ({\it dark}) and 2$\sigma$ ({\it light}) 
in the plane $\varpi_0$-S for WMAP+ACBAR ({\it top panel}) and CMBpol ({\it bottom panel}).}
 \label{fig2}
\end{center}
\end{figure}

% Gravitational Waves

The E- to B-conversion is on an angular scale where it is not
contaminated by primordial gravitational wave signal in the B-modes, which are relevant at
larger angular scales, if at all. And although the implicitly assumed theory of
gravitation should introduce new degrees of freedom, the scalar-vector-tensor
decomposition of perturbations in linear theory ensures us that no further
sources of B-mode polarization should arise. We further caution that viable
models must satisfy $|\varpi| \lesssim 10^{-5}$ within the Solar System, with a
transition taking place near the outskirts of the galaxy.  Rather than implying a
scale-dependence for post-GR effects,  this suggests that a viable model for
$\varpi$ must display a nonlocal or environmental dependence on the density
field, with $\varpi$ vanishing within a few tens of kpc of a galactic core.
CMB photons are weakly lensed by Mpc-scale density pertubations, but
should not experience post-GR effects while passing so near to
galactic cores. On the celestial sphere, kpc radii subtend angular scales well
below the angular scales of interest for next-generation polarization experiments.
Thus, we tentatively ignore the position-dependence of the post-GR effects introduced by
Eq.~\ref{eqn:varpi}.

In the case of temperature anisotropies, at small angular scales where the
lensing effect is present, confusion from other secondary signals, most notably
the Sunyaev-Zel'dovich effect \cite{Cooray:2000ge,Komatsu:2002wc} and clustering
of unresolved extragalactic point sources \cite{Serra:2008ge}, must also be
considered. When fitting to WMAP and ACBAR
data, we take into account the contribution from
clustered point sources on the angular power spectrum in order to avoid a bias in
the determination of the cosmological parameters. We do this by writing the total
CMB anisotropy spectrum as $C_\ell^{tot}= C_\ell^{CMB}+C_\ell^{PS}+C_\ell^{SZ}$
and allowing the amplitudes of both the SZ contribution $(A_{sz})$ and clustered
point sources (with varying amplitudes for the two different experiments
\cite{Serra:2008ge}) to vary as free parameters when fitting for the combined
cosmological and post-GR parameters.

We make use of the publicly available Markov Chain Monte Carlo (MCMC) package
CosmoMC \cite{Lewis:2002ah} with a convergence diagnostic based on the Gelman and
Rubin statistic \cite{gelman} to model fit the data. The post-GR parameters are
allowed to take values $-5 <\varpi_0 <10$ and $0<S<10$.  In addition, we implement
the flat $\Lambda$CDM cosmological model with six standard parameters: baryon
density $\Omega_b h^2$; dark matter density $\Omega_c h^2$; reionization optical
depth $\tau$; ratio of the sound horizon to the angular diameter distance at the
decoupling measured by $\theta$; amplitude of the curvature perturbation $A_s$
(with flat prior on $log(A_s)$; spectral index $n_s$. These two last parameters
are defined with respect to a pivot scale at  $0.002$ h/Mpc, as in
Ref.~\cite{Dunkley:2008ie}.

\smallskip
\noindent \emph{Results---}We first use the combination of WMAP 5-year
\cite{Komatsu:2008hk} and ACBAR data \cite {Reichardt:2008ay} (both temperature
and temperature-polarization cross-correlation). To avoid complications due to
overlapping of the datasets, we use WMAP data out to $\ell<900$ and then ACBAR
data in the range $900<\ell<2000$.   The constraint on $\varpi_0$ is $\varpi_0 = 1.67^{+3.07}_{-1.87}$ 
at the 2$\sigma$ confidence level, but $S$ remains unconstrained.
As shown in the upper panel of Fig.~\ref{fig2}, there is a 
clear correlation between $S$ and $\varpi_0$: when $S$ goes to $0$ only 
very small values of $\varpi_0$ are allowed and when $S\sim5$, 
values of $\varpi_0\sim6$ are allowed at the $2\sigma$ confidence level. 

By including the post-GR parameterization, we find that cosmological parameters from WMAP+ACBAR
change by less than 1$\sigma$; for example, with post-GR effects,
$\sigma_8=0.814\pm0.044$ and $n_s=0.954\pm0.014$ while $\sigma_8=0.803\pm0.034$
and $n_s=0.964\pm0.014$ \cite{Komatsu:2008hk} without post-GR effects.

To study the extent to which future CMB data improve these constraints, we create
mock datasets for both Planck and CMBpol. For Planck we create a mock temperature and polarization dataset
with noise properties consistent with a combination of Planck at $100$, $143$,
and $217$ GHz channels of HFI \cite{bluebook}. 
We assume the best fit WMAP5$+$ACBAR parameters without modified
gravity \cite{Komatsu:2008hk} as the underlying cosmological model. 
We use the full-sky likelihood function given in
Ref.~\cite{Lewis:2005tp} when fitting the data. The upper and lower limits with
Planck don't show improvement compared to the case with WMAP and ACBAR; $S$ is still unconstrained  
and $\varpi_0<5.08$ (2$\sigma$), mostly due to degeneracies
with other cosmological parameters. While we include polarization information,
Planck does not probe the lensed B-mode spectrum with adequate signal-to-noise
ratio, as seen in the middle panel of Fig.~\ref{fig1}.

%\begin{figure}[!t]
%\begin{center}
%\includegraphics[scale=1.0,angle=0,width=8cm]{fig6new.eps}
%\end{center}
%\caption{Two-dimensional contours at $1\sigma$ and $2\sigma$ in the plane $\varpi_0$-S for
%  WMAP+ACBAR and CMBpol .}
%\end{figure}

To study how improved polarization measurements, and thereby a measurement of the 
lensing potential power spectrum, improve the parameter constraints, we also make mock datasets for
CMBpol using $70$ GHz to $220$ GHz for the 2-meter version of the EPIC concept
study \cite{Bock:2008ww}.  We also make a mock dataset of the lensing potential power spectrum
under the same cosmological model by creating the noise spectrum for the lensing
construction with using same experimental noise as above concept with the reconstruction calculated using
quadratic statistics \cite{Hu:2001kj}. To avoid any biases from non-linearities,
we consider only multipoles out to $l<700$ probed by CMBpol. 
The projected upper limit with CMBpol is $\varpi_0<0.30$ (2$\sigma$), showing
significant improvement compared to the present-day CMB constraints; the parameter 
$S$ remains unconstrained as we find the same strong correlation with the amplitude $\varpi_0$.

Previous studies have shown that the combination of Planck and a probe of the
large-scale structure weak lensing such as from NASA/DOE JDEM or the ESA-based
Euclid can improve constraints of modified gravity. With $S=3$ fixed, the Planck
and future weak lensing combination constrains  $\varpi_0 =
-0.07^{+0.13}_{-0.16}$ at the 95\% confidence level \cite{Daniel}. In comparison with an
experiment such as CMBpol using lensing information from B-modes,
we have considered the constraints on both the amplitude and the redshift-dependence of the
post-GR effects. If we fix $S=3$, then for CMBpol we find $\varpi_0 < 0.11$
(2$\sigma$), which is comparable to the projections for Planck and a half-sky space-based weak lensing
survey with Euclid. While competitive, the advantage with the CMB constraint is that it comes
from a single dataset and avoids issues related to systematics that can impact
weak lensing observations.

%Here we have considered a redshift dependent modification to post-GR
%parameterization improving previous constraints that assumed an evolution
%following the ratio of dark energy to matter density with $S=3$. In an upcoming
%paper we will generalize the treatment to a redshift and scale-dependent
%modification to gravity.

%%%%%%%%%%%%%%%%%%%%%%%%%%%%%%%%%%%%%%%%%%%%%%%%%%%%%%%%%%%%%%%%%%%%%%%%%
%%%%%%%%%%%%%%%%%%%%%%%%%%%%%%%%%%%%%%%%%%%%%%%%%%%%%%%%%%%%%%%%%%%%%%%%%
\smallskip
AC and PS acknowledge support from NSF CAREER AST-0645427. 
PS thanks Alex Amblard for useful discussions.
RC and SD are supported by NSF AST-0349213 at Dartmouth.

%%%%%%%%%%%%%%%%%%%%%%%%%%%%%%%%%%%%%%%%%%%%%%%%%%%%%%%%%%%%%%%%%%%%%%%%%
%%%%%%%%%%%%%%%%%%%%%%%%%%%%%%%%%%%%%%%%%%%%%%%%%%%%%%%%%%%%%%%%%%%%%%%%%

\end{document}